\documentclass{llncs}

\usepackage{amssymb}
\usepackage{amsmath}
\usepackage{epic}
\usepackage{eepic}
\usepackage{graphicx}
\DeclareSymbolFontAlphabet{\mathbb}{AMSb}
\usepackage{times}
\usepackage{amsfonts}
\usepackage{eucal}
\newtheorem{thm}{Theorem}
\newtheorem{lm}[thm]{Lemma}

\begin{document}

\title{Capacity as a Fundamental Metric for Mechanism Design in the Information Economy}

\author{ Sudhir Kumar Singh \and  Vwani P. Roychowdhury }
\institute{ Department of Electrical Engineering, University of California, Los Angeles, CA 90095.\\
{\it \{sudhir,vwani\} @ ee.ucla.edu}}

\maketitle

\begin{abstract}
The auction theory literature has so far focused mostly on the design of mechanisms that
takes the revenue or the  efficiency as a yardstick.
 However,  scenarios where the {\it capacity},
 which we define as \textit{``the number of bidders the auctioneer
wants to have a positive probability of getting the item''},  is a fundamental concern are
ubiquitous in the information economy. For instance, in  sponsored search auctions (SSA's)
or in online ad-exchanges, the true value of an ad-slot for an advertiser is inherently
derived from the conversion-rate, which in turn depends on whether the advertiser actually
obtained the ad-slot or not; thus, unless the capacity of the underlying auction is large,
key parameters, such as true valuations and advertiser-specific conversion rates, will
remain unknown or uncertain leading to inherent inefficiencies in the system. In general,
the same holds true for all information goods/digital goods. We initiate a study of
mechanisms, which take capacity as a yardstick, in addition to revenue/efficiency. We show
that in the case of a single indivisible item one simple way to incorporate capacity
constraints is via designing mechanisms to sell probability distributions, and that under
certain conditions, such optimal probability distributions could be identified using a
Linear programming approach. We define a quantity called {\it price of capacity} to
capture the tradeoff between capacity and revenue/efficiency. We also  study the case of
sponsored search auctions. Finally, we discuss how general such an approach via probability
spikes can be made, and potential directions for future investigations.   
\end{abstract}

\section{Introduction}
\begin{center}
{\it ``The tension between giving away your information- to let people know what you have to offer- and charging 
them for it to recover your costs is a fundamental problem in the information economy.''}
\end{center}

 --- Carl Shapiro and Hal R. Varian in 
Information Rules: A Strategic Guide to the Network Economy, Harvard Business School Press (1998).  \\

The advent of the Internet as a big playground for resource sharing among selfish agents with diverse interests, 
and the emergence of Web as a giant platform for hosting information has raised a plethora of opportunities 
for commerce, as well as, a plenty of new design, pricing and complexity problems. 
One good example of a multi-billion dollar industry evolved as a consequence of Web is the sponsored search advertising (SSA),  
making fortunes for Internet Search giants such as Google and Yahoo!, 
and has got tremendous attention in academia 
recently, due to various interesting research problems originated as 
a result of this continuously growing industry\cite{AGM06,BCPP06,EOS05,Lah06,LP07,MSVV05,PO06,SBRR07,cap-med,diverse,Var06,WVLL07}.
One of the most important concern for such an industry or in general for the information economy is the {\it ``pricing problem''}.
For example, for the goods like an ad-slot in the SSA which has no intrinsic value and is perishable, 
it is not clear what price should it be sold for. Similarly,
 for a digital good, where the  cost of reproduction is negligible, 
the standard way of pricing based on the production cost does not work.  
Therefore, 
auctions are becoming a popular pricing mechanism in electronic commerce as 
they automatically adjust prices to market conditions, and specifically prices gets adjusted according 
to its value to the consumers rather than to the production costs. 
Auction theory has a pretty impressive 
literature \cite{krishna,milgrom}- from the lovely Vickrey auction to the 
sponsored search auctions\cite{EOS05,Var06} and the auctions of digital goods\cite{GHW01,GHKSW06}. 
The literature has so far focused mostly on the design of mechanisms 
that takes the revenue or the efficiency as a yardstick. 
This is perfectly logical as these are two very important metrics from the viewpoint of the seller/auctioneer and 
the society respectively. However, the scenarios where the {\it capacity}, 
 which we define as \textit{``the number of bidders the auctioneer 
wants to have a positive probability of getting the item''},  is a fundamental concern are ubiquitous 
in the information economy\footnote{Clearly, this capacity can be increased 
indefinitely albeit sometimes at the cost of revenue.}. 
For instance, in the sponsored search auctions or in online ad-exchanges, the true value of an ad-slot for an advertiser is 
inherently derived from the conversion-rate, which in turn depends on whether the advertiser actually 
obtained the ad-slot or not, which in turn depends on the capacity. In general,
the same holds true for all information goods/digital goods.  
 
In the present paper, our goal is to first motivate {\it capacity} as a fundamental metric in designing auctions in the 
information economy and then to initiate study of such a design framework for some simple and interesting scenarios. In Section \ref{mot},
we motivate the capacity as a fundamental and interesting additional metric on the top of revenue/efficiency for 
mechanism design. In Section \ref{single}, we start with the {\it capacity } constrained framework for selling an 
indivisible item. We propose a simple way 
to incorporate capacity constraints via designing mechanisms 
to sell probability distributions.
We show that such optimal probability distributions could be identified
using a Linear programming approach, objective being revenue, efficiency or a related function. 
Further, we define a quantity called {\it price of capacity} to capture the tradeoff 
between capacity and revenue/efficiency and derive upper bounds on it.
In Section \ref{ssa}, we discuss the case of sponsored search auctions and also note that 
the auctioneer controlled probability spikes based auctions suggests a new model for sponsored search advertising,
where a click is sold directly and not indirectly via allocating impressions. In Section \ref{conc}, we conclude with a list of 
research directions for future work, inspired by the present paper.  

\section{Motivation: the need for an additional metric}
\label{mot}

\paragraph{ {\bf Experience goods:}} 
A bidder might not know her true valuation for the item unless she acquires it sometimes
meaning that the true value is inherently derived from the actual acquisition of the item.  
Such a good is called an {\it experience good}\cite{Nelson70,inforules,infoeco}.  
Experience goods are ubiquitous in the information economy as clearly all information goods are experience goods\cite{inforules,infoeco}.  
Sometimes, a particular good might also act as an experience for another good. For example, a particular song from a singer might act as an experience 
for another song of that singer. Therefore, in the auction of an experience good, the values for some of the bidders can be
known to them as they might have 
experienced it from an earlier purchase of this or a related item, and for the other bidders the values are still unknown and they have to simply
guess this value if they are participating in the auction.
For the second kind of bidders, even if their values might turn out to be pretty high, their guesses might not be high enough to 
actually acquire the item when revenue/efficiency is the only goal. 
Moreover, they might be {\it loss-averse}\cite{TK91} and would not bid a high value at all, due to the potential risk involved
if the item does not turn out to be of high value to them. Therefore, it is important that such bidders be given a 
chance to acquire the item, and consequently capacity becomes a fundamental metric in designing the mechanisms
to achieve this goal.    
 
\paragraph{ {\bf Two-fold exploration in sponsored search auctions:}} 
First, the ad-slots in the SSA are necessarily experience goods 
for a new advertiser and the estimates for the advertisers getting lower ranked slots is
also generally poor as they hardly get any clicks.
The value of an ad-slot is
 derived from the clicks themselves (i.e. rate of conversion or purchase given a
click), and therefore, unless the bidder actually obtains a slot and receives user clicks, there
is essentially no means for her to estimate her true value for the associated keyword.  

Second, even if all the true valuations are known to the corresponding bidders,
for each bidder the SSA involves a parameter called {\it quality score} of the bidder 
which is defined as the expected clickability of the bidder for the associated keyword if she obtains a slot. 
This parameter is also not known a priori  and the auctioneer needs to estimate it. 
Certainly, a
model that automatically allows one to estimate these key parameters (i.e. Click-Through-Rates and true
values) is desirable. Indeed, some mechanisms to incorporate explorations for estimating such important parameters
has started to appear in literature\cite{SBRR07,WVLL07}. 
Capacity as an additional metric can provide a generic framework for designing such exploration based mechanisms.  

\paragraph{ {\bf Avoiding over-exposure in online advertising: }} Typically, the online ad-exchanges such as Right Media or DoubleClick
convince their advertisers that their ads will not be over exposed to users. One way of avoiding such an over-exposure could be
via increasing capacity.    

\paragraph{ {\bf Uncertainty and switching costs:}} Let us consider a production company $H$ buying raw materials (multiple units of a good) 
from providers $A, B$ and $C$ via a reverse-auction and $H$ is uncertain about the time these providers might 
take to deliver the raw material to $H$. The providers are very likely to lie about the delivery times and it is hard 
to incorporate delivery times in designing the auction.
If the goal of the auctioneer (i.e. the company $H$) is cost minimization, it will buy the raw material from the 
provider with the minimum ask (assume that the ask of $B$ is smaller than that of $A$ which is smaller than that of $C$). 
Now if the provider $B$ lied about the delivery time at least for a 
significant fraction of the total required units, $H$'s production gets delayed. If $H$ wants to switch to some other 
provider, run another auction 
and  buys from $A$, now it will buy at a higher cost and their is still a delay in $H$'s production 
as $A$ will take its time in delivery too.
The time taken by $A$ could actually be smaller than that of $B$'s for the remaining units,
however, still there is a delay. Moreover, such delay might persist further
as $A$ could also lie about its delivery time. 
It might have been better if $H$ would buy not only from $B$ to start with and give $A,C$ a chance as well. 
Therefore, one way to reduce such delay times could be via increasing {\it capacity} as per our definition.

\section{Selling an indivisible item}
\label{single}

\subsection{The model to sell via probability spikes}
There is a single {\it indivisible} item for sale.
There are $N$ bidders interested in the item.
The bidder $i$ has a value $v_i$ for this item\footnote{ 
Note that, as discussed earlier in Section \ref{mot}, 
for some bidders $v_i$'s are the actual true values while for some others these are just crude estimates/guesses.}.  
The item is sold via an auction on an experiment
designed by the seller/auctioneer. The experiment 
has $M$ outcomes $(O_1,O_2,\dots,O_M)$
with associated probabilities 
$(p_1,p_2,\dots,p_M)$, where $\sum_{i=1}^M p_i=1, p_i \geq 0 \forall i$. 
Therefore, the item 
is essentially sold via an auction of the probability 
spikes  $(p_1,p_2,\dots,p_M)$ wherein 
the auctioneer can choose these probability 
spikes in advance or adaptively based on bidders' reports 
so as to achieve some defined goal such 
as maximizing her profit or efficiency or to accommodate a wider pool of bidders. 
Bidders bid on the experiment by reporting bids $b_i$'s to indicate 
their respective values of the item. 
{\it At most one probability spike is assigned to each bidder.} 
Thus there are effectively two steps in this auction model.
\begin{itemize}
\item Stage 1 ({\em commit/compete}): The bidders 
report their bids $b_i$'s and by way of using some mechanism, 
the auctioneer assigns the probability spikes to them and 
decides corresponding payments to be made by them.
Let us call a bidder a {\it prospective winner} if she 
was assigned one of the probability spikes. 
\item Stage 2 ({\em win or lose}):  
The experiment is performed.
If the outcome of the experiment is $O_j$, then
the {\it prospective winner} assigned to the 
spike $p_j$ is declared the {\it winner}, and is given the 
item.   
 \end{itemize}    
Further, the auctioneer could choose various payment schemes such as -
\begin{itemize}
\item {\em Betting:} Every {\it prospective winner} is charged its payment 
decided in {\it compete/commit} stage irrespective of whether she will 
be a {\it winner} or not.  
\item {\em Pay-per-acquisition}:  
A bidder is charged the amount decided in {\it compete stage} 
only when she is a {\it winner} i.e. only 
when she actually acquires the item.  
\end{itemize}
The above model can also be interpreted as selling of a single {\it divisible} item
in terms of specified fractional bundles, the bundles corresponding to
the probability spikes. 

\subsection{Mechanisms to sell probability spikes}
Without loss of generality, let us assume that 
$p_1 \geq p_2 \dots \geq p_M \geq 0$.
Further, for notational simplicity let $\mathcal{M}=\{1,2,\dots,M\}, \mathcal{N}=\{1,2,\dots,N\}$.
Let $\sigma: \mathcal{N} \longrightarrow \mathcal{N} $ be the allocation
rule and $h: \mathcal{N} \longrightarrow R_+$ be the payment rule decided 
in {\it compete stage}.
Thus, for $j \in \mathcal{M}$, the spike $j$ is assigned to the bidder $\sigma(j)$ and for $j \in \mathcal{N}-\mathcal{M}$, the bidder $\sigma(j)$  
is not assigned any spike. Further, for $j \in \mathcal{M}$, $h_j$ is the expected payment to be made by the bidder $\sigma(j)$ and $h_j=0$ otherwise.
Therefore, the expected utility of the bidder assigned to 
spike $j$ is given by 
$u_{\sigma(j)} = p_j v_{\sigma(j)} - h_j$ for $j \in \mathcal{M}$ and is zero otherwise.
For the sake of simplicity, let $p_1 > p_2 \dots > p_M > 0$, 
then the famous {\em VCG} mechanism ranks the bidders by their bids 
(and true values $v_i$'s as being a truthful mechanism)  
and charges them their respective opportunity costs. 
That is, $\sigma(j)$ is the bidder with the $j$th maximum bid
and 
$h_j = \sum_{i=j}^{M-1} (p_i - p_{i+1}) v_{\sigma(i+1)} + p_M  v_{\sigma(M+1)} $ for $j \in \mathcal{M}$ and zero otherwise.    
If the payment is done via the {\it betting} model then 
this is the amount charged to bidder $j$ in the {\it compete/commit} stage. 
If the payment is done via {\it pay-per-acquisition} and $j$ is the {\it winner} then
she is charged an amount $\frac{h_j}{p_j}$, and therefore, her expected payment is 
still $h_j$ as $p_j$ is the probability that she wins.  
Therefore, the auctioneer's revenue is 
\begin{align}
R_{VCG} = \sum_{i=1}^{M-1} (p_i - p_{i+1}) i v_{\sigma(i+1)} + p_M M v_{\sigma(M+1)}.    
\end{align} 
Let $\theta_j = p_j - p_{j+1} ; j=1,2,\dots,M-1 \& \theta_M = p_M$ 
be the spike gaps, then
$p_j = \sum_{i=j}^M \theta_i$.
The condition  $\sum_{i=1}^M p_i=1$ translates to
$\sum_{i=1}^M i \theta_i = 1$
and $p_1 \geq p_2 \dots \geq p_M \geq 0$
translates to 
$ \theta_j \geq 0 \forall j =1,2,\dots,M.$ Therefore,
\begin{align}
R_{VCG} = \sum_{i=1}^M \theta_i i d_i  \textrm{ where  } d_i= v_{\sigma(i+1)}    
\end{align} 
and clearly $d_i \geq d_{i+1}$ as {\it VCG } ranks by true values. 
\begin{lm}
\label{vcgrev}
The revenue of the auctioneer in the {\em VCG} mechanism for selling probability spikes $(p_1,p_2,\dots,p_M)$ 
can be expressed as  $R_{VCG} = \sum_{i=1}^M \theta_i i d_i$ where $\theta_i =p_i -p_{i+1}$ for all 
$i=1,2,\dots,M-1$, $\theta_M=p_M$ and $d_i \geq d_{i+1}$ for all $i=1,2,\dots,M$. 
\end{lm}
Further, the efficiency for the {\em VCG} mechanism is 
\begin{align}
E_{VCG} = \sum_{j=1}^M p_j v_{\sigma(j)} = \sum_{j=1}^M (\sum_{i=j}^M \theta_i) v_{\sigma(j)}  \nonumber \\
   = \sum_{i=1}^M (\sum_{j=1}^i v_{\sigma(j)}) \theta_i = \sum_{i=1}^M \theta_i i d_i \\
\textrm{where  }  d_i = \frac{1}{i} \sum_{j=1}^i  v_{\sigma(j)} .
\end{align}
Further, we have 
\begin{align*}
d_i - d_{i+1} & =  \frac{1}{i} \sum_{j=1}^i v_{\sigma(j)} - \frac{1}{(i+1)} \sum_{j=1}^{i+1} v_{\sigma(j)} = \frac{1}{i(i+1)} \left\{\sum_{j=1}^i(i+1) v_{\sigma(j)} - \sum_{j=1}^{i+1}i  v_{\sigma(j)}\right\} \\
             & = \frac{1}{i(i+1)} \left\{\sum_{j=1}^i v_{\sigma(j)} - i  v_{\sigma(i+1)}\right\}  = \frac{1}{i(i+1)} \left\{\sum_{j=1}^i( v_{\sigma(j)} - v_{\sigma(i+1)}) \right\} \\
             & \geq 0  \textrm{ as }  v_{\sigma(j)} \geq v_{\sigma(i+1)}\forall j=1,2,\dots, i .
\end{align*}
\begin{lm}
\label{vcgeff}
The efficiency in the {\em VCG} mechanism for selling probability spikes $(p_1,p_2,\dots,p_M)$ 
can be expressed as  $E_{VCG} = \sum_{i=1}^M \theta_i i d_i$ where $\theta_i =p_i -p_{i+1}$ for all 
$i=1,2,\dots,M-1$, $\theta_M=p_M$ and $d_i \geq d_{i+1}$ for all $i=1,2,\dots,M$. 
\end{lm}

\begin{definition}
We say that a linear function $H$ of spike-gaps $\theta_j$'s is {\em gap-wise monotone}  if 
$H = \sum_{j=1}^M \theta_j j d_j$, where $d_j$'s do not depend on gaps $\theta_j$'s and $d_j \geq d_{j+1}$ for all $j=1,2,\dots,M$. 
\end{definition}
\begin{definition}
A mechanism for selling probability spikes is called {\em gap-wise monotone}  if 
the revenue of the auctioneer at the prescribed equilibrium point is  {\it gap-wise monotone} and is called  {\em  strongly gap-wise monotone} if 
the social value (i.e. efficiency) at the prescribed equilibrium point is  {\it gap-wise monotone} as well.   
\end{definition}
Therefore, from Lemma \ref{vcgrev} and Lemma \ref{vcgeff} we obtain the following theorem.
\begin{thm}
\label{vcg}
The {\em VCG} mechanism for selling probability spikes is {\it strongly gap-wise monotone}. 
\end{thm}

Define, 
$u_i^{*}(h) = \max\{ \max_{ j \in \mathcal{M}} (p_j v_i - h_j) , 0\}$. 
\begin{definition}
{\bf Walrasian Equilibrium:} Let $\sigma$ be an allocation and $h$ be a payment rule, then 
$(\sigma, h)$ is called a Walrasian equilibrium if for all $j \in \mathcal{N}$,
$u_{\sigma(j)} = u_{\sigma(j)}^{*}(h)$.
\end{definition}
Following \cite{SS72,DGS86,BO06}, it is not hard to establish the following lemma.
\begin{lm}
Let $\sigma$ be an allocation and $h$ be a payment rule, then 
$(\sigma, h)$ is a Walrasian equilibrium iff it is efficient.
\end{lm}
Therefore, at a Walrasian equilibrium, bidders are ranked according to their values
and efficiency can be written as in the case of VCG i.e $\sum_{j=1}^M p_j v_{\sigma(j)}$
where $  v_{\sigma(j)} \geq  v_{\sigma(j+1)}$ for all $ j =1,2,\dots,M$.
\begin{thm}
Let $(\sigma, h)$ be a Walrasian equilibrium for selling probability spikes then the efficiency at this 
equilibrium is gap-wise monotone. 
\end{thm}
This means that optimal efficiency is always gap-wise monotone. 
Further, the optimal omniscient auction (i.e. when the auctioneer knows everyone's true value $v_i$'s)
extracts a revenue equal to $\sum_{j=1}^M p_j v_{\sigma(j)}$, where $v_{\sigma(j)} \geq v_{\sigma(j+1)}$ for all $ j =1,2,\dots,M-1$ .
Therefore, the optimal revenue of omniscient auction is also gap-wise monotone. 
 
\subsection{A Generic Framework for Selecting Optimal Spikes}
\label{genopt}
In this section, we develop a Linear Programming approach to identify optimal probability spikes 
subject to the capacity constraints in terms of spikes gaps,  
where the objective is a gap-wise monotone function.
For such functions, it is simpler to put the constraints in terms of spike-gaps 
than in terms of spikes themselves, however,  it won't be hard to see 
that a similar approach can also be developed if we put the constraints in terms of the spikes, as well as, in the case 
of functions more general than the gap-wise monotone. For the sake of simplicity we omit any such details.  

Let $H$ be a {\it gap-wise monotone} function and $\{\epsilon_j\}_{j=1}^M$ be a generic set of parameters
with the property that 
\begin{align}
\sum_{i=1}^M i \epsilon_i \leq 1. \label{epsilonjs0}\\
\epsilon_j \geq 0 ;  j=1,2,\dots, M \label{epsilonjs}
\end{align}
and let us consider 
the following Linear Programming Problem in variables $\theta_j$'s,   
\begin{align}
\textrm{ {\em Max} } & H = \sum_{j=1}^M \theta_j j d_j \nonumber\\
\textrm{ s.t. } & \sum_{i=1}^M i \theta_i = 1 \\
& \theta_j \geq \epsilon_j ; j=1,2,\dots,M \label{gap-ep}  
\end{align}  
The dual problem is 
\begin{align}
\textrm{ {\em Min} } &  x_0 - \sum_{j=1}^M \epsilon_j x_j \nonumber\\
\textrm{ s.t. } & x_j \geq 0 ; j=1,2,\dots,M\\
& -jd_j+jx_0 -x_j=0 ; j=1,2,\dots,M \nonumber 
\end{align} 
and the {\em KKT} conditions are 
\begin{align}
\sum_{i=1}^M i \theta_i = 1 \nonumber \\
 \theta_j \geq \epsilon_j ; j=1,2,\dots,M \nonumber \\
 x_j \geq 0 ; j=1,2,\dots,M \\
 -jd_j+jx_0 -x_j=0 ; j=1,2,\dots,M \nonumber \\
x_j (\epsilon_j-\theta_j) =0 ; j=1,2,\dots,M \nonumber
\end{align}
and therefore an optimal solution is 
\begin{displaymath}
\begin{array}{l}
x_0^* = d_1 \\
x_j^*=j(d_1 - d_j) \\
\end{array}
\end{displaymath}
\begin{align} 
\label{optsol}
\theta_j^* = \epsilon_j \forall j=2,\dots, M  \nonumber\\
\theta_1^* = 1 - \sum_{i=2}^M i \epsilon_i  
\end{align}
as it can be checked to satisfy the {\em KKT} conditions.   
The optimal value is 
\begin{align}
H^{OPT}(\{\epsilon_j\}) = d_1 - \sum_{j=2}^M j \epsilon_j (d_1-d_j) \label{H-opt}.
\end{align}
Clearly, the maximum of the optimal solution 
$H^{OPT}(\epsilon_1,\epsilon_2,\dots,\epsilon_M)$
over parameters $\{\epsilon_j\}$'s is attained when $\epsilon_j=0$ for all $j=1,2,\dots,M$ and in that case
\begin{align}
H^{OPT} = d_1.
\end{align}
Now,  note that the primal optimal variables $\theta_j$'s do not 
depend on the quantities $d_j$'s at all.
Therefore, so long as $H$ is {\it gap-wise monotone}, 
the optimal solution to the primal remains the same as in equation \ref{optsol}. 
It is quite intuitive as the best possible spike allowed by the capacity constraints
is assigned to the best possible bidder (i.e. $\sigma(1)$),
and all other spikes are the minimum possible as per the constraints. 
\begin{thm}
Let $H = \sum_{j=1}^M \theta_j j d_j$ be a gap-wise monotone function and the spike-gaps $\theta_j$'s satisfy conditions \ref{gap-ep}, \ref{epsilonjs0}
and  \ref{epsilonjs}, then the optimal choice of spike-gaps are given by equation \ref{optsol}
and the optimal value of $H$ is given by equation \ref{H-opt} .  
\end{thm}
\subsection{The Price of Capacity}
Given parameters $\{\epsilon_j\}$, let us define 
the {\it capacity} as
\begin{align}
\kappa(\{\epsilon_j\}) = \max_{j} \{ j: \epsilon_j > 0\}.    
\end{align}
Now consider the parameters $\{\tilde{\epsilon}_j\}$ satisfying the properties \ref{epsilonjs0} and \ref{epsilonjs} such that 
$\kappa(\{\tilde{\epsilon}_j\}) = \kappa(\{\epsilon_j\})+1$.
Given a  gap-wise monotone function $H = \sum_{j=1}^M \theta_j j d_j$, we claim that such $\tilde{\epsilon}_j$'s
satisfying properties \ref{epsilonjs0} and \ref{epsilonjs} can always be obtained from $\epsilon_j$'s
such that  $H^{OPT}(\{\tilde{\epsilon}_j\}) \geq H^{OPT}(\{\epsilon_j\})$
as long as  $ H^{OPT}(\{\epsilon_j\}) < H^{OPT}$, meaning that the {\it capacity}
can always be increased without any loss in optimal value as long as we do not shoot over the absolute optimum  $H^{OPT}$.
We have,
\begin{displaymath}
\begin{array}{l}
H^{OPT}(\{\tilde{\epsilon}_j\}) - H^{OPT}(\{\epsilon_j\}) 
= \sum_{j=2}^{\kappa(\{\epsilon_j\})} j \epsilon_j (d_1-d_j) - \sum_{j=2}^{\kappa(\{\epsilon_j\})+1} j \tilde{\epsilon}_j (d_1-d_j) \\
= \sum_{j=2}^{\kappa(\{\epsilon_j\})} j(\epsilon_j-\tilde{\epsilon}_j) (d_1-d_j)
- (\kappa(\{\epsilon_j\})+1) \tilde{\epsilon}_{\kappa(\{\epsilon_j\})+1}  (d_1-d_{\kappa(\{\epsilon_j\})+1}).
\end{array}
\end{displaymath}
Now we can always choose  $\tilde{\epsilon}_j$'s satisfying properties \ref{epsilonjs0} and  \ref{epsilonjs} 
by taking suitable
 $\tilde{\epsilon}_j \leq \epsilon_j ; j=1,\dots,\kappa(\{\epsilon_j\})-1 $, 
$\tilde{\epsilon}_{\kappa(\{\epsilon_j\})} < \epsilon_{\kappa(\{\epsilon_j\})}$ and
$\tilde{\epsilon}_{\kappa(\{\epsilon_j\})+1} \leq \frac{ \sum_{j=2}^{\kappa(\{\epsilon_j\})} j(\epsilon_j-\tilde{\epsilon}_j) (d_1-d_j)}{(\kappa(\{\epsilon_j\})+1)  (d_1-d_{\kappa(\{\epsilon_j\})+1})}$. 
In particular, taking $\tilde{\epsilon}_j = \epsilon_j ; j=1,\dots,\kappa(\{\epsilon_j\})-1 $, 
$\tilde{\epsilon}_{\kappa(\{\epsilon_j\})} < \epsilon_{\kappa(\{\epsilon_j\})}$ and
$\tilde{\epsilon}_{\kappa(\{\epsilon_j\})+1} \leq \min \left\{ \frac{ \sum_{j=2}^{\kappa(\{\epsilon_j\})} j(\epsilon_j-\tilde{\epsilon}_j) (d_1-d_j)}{(\kappa(\{\epsilon_j\})+1)  (d_1-d_{\kappa(\{\epsilon_j\})+1})}, 
\frac{ \kappa(\{\epsilon_j\}) (\epsilon_{\kappa(\{\epsilon_j\})}) - \tilde{\epsilon}_{\kappa(\{\epsilon_j\})}}
{(\kappa(\{\epsilon_j\})+1)} \right\}$ does the job. 
An interesting case to consider is when $\epsilon_j=\epsilon > 0$ for all $j \leq m$ and  $\epsilon_j=0$ otherwise. 
And in this case we can increase the capacity without loss in optimal value by taking $\tilde{\epsilon}_j=\tilde{\epsilon}$ and  $\tilde{\epsilon}_j=0$ otherwise, where 
\begin{align}
\tilde{\epsilon} \leq  \min \left\{ \frac{2}{(m+1)(m+2)} , \frac{\sum_{j=2}^m j (d_1-d_j)}{\sum_{j=2}^{m+1}j (d_1-d_j)}\epsilon\right\}.
\end{align}
Now let us define
\begin{align}
a:=\min_j {j: d_1 > d_j} \nonumber
\end{align}
then it is clear that 
\begin{align}
H^{OPT}(\{\epsilon_j\}) & = H^{OPT} \textrm{ whenever } \kappa(\{\epsilon_j\}) < a \nonumber \\
& < H^{OPT} \textrm{ whenever } \kappa(\{\epsilon_j\}) \geq a \nonumber
\end{align}
and therefore, as long as $\kappa(\{\epsilon_j\}) \geq a$ capacity can always be increased without any loss in the optimal value 
as discussed above, however there is a strict decrease in the optimal value if we wish to increase capacity from $a-1$ to $a$.
We can naturally define a parameter which we call {\it price of capacity} as follows:
\begin{align}
\nu(\{d_j\}) & := \max_{\{\epsilon_j\}:\kappa(\{\epsilon_j\})=a} \left( \frac{H^{OPT}}{H^{OPT}(\{\epsilon_j\})}\right) \\
& =  \max_{\{\epsilon_j\}:\kappa(\{\epsilon_j\})=a} \left(\frac{d_1}{d_1-a\epsilon_a(d_1-d_a)}\right) \nonumber\\
& \leq \frac{d_1}{d_a} = \frac{d_{a-1}}{d_a}
\end{align}
Thus, {\it price of capacity} is the worst possible loss in optimal value while increasing capacity from $a-1$ to $a$.  
Again let us consider the case when all non-zero $\epsilon_j =\epsilon$, then 
\begin{align}
\nu(\{d_j\}) & = \max_{0 < \epsilon \leq \frac{2}{a(a+1)}} \left(\frac{d_1}{d_1-a\epsilon(d_1-d_a)}\right) = \frac{d_1}{d_1 -\frac{2}{a+1} (d_1 -d_a)} \nonumber \\
& = \frac{(a+1)}{(a-1)+2\left(\frac{d_a}{d_1}\right)} \leq 1 + \frac{2}{a-1} \leq 3
\end{align}

Often our goal will be to maximize efficiency or revenue subject to the capacity constraints, and consequently such a 
loss may not be considered good. Therefore, its really a price that we are paying for increasing capacity.

\section{Sponsored Search Auctions}
\label{ssa}
As we discussed in the Section \ref{mot}, one nice motivation for the study of capacity as a metric for mechanism design 
comes from the sponsored search advertising.
We first describe the formal SSA model. 
Formally, in the current models, there are $K$ slots to be allocated among  $N$ ($\geq K$)
bidders (i.e. the advertisers). A bidder $i$ has a true valuation $v_i$ (known only to the
bidder $i$) for the specific keyword and she bids $b_i$. The expected {\it click through
rate} (CTR) of an ad put by bidder $i$ when allocated slot $j$ has the form $CTR_{i,j}=\gamma_j e_i$
i.e. separable in to a position effect and an advertiser effect. $\gamma_j$'s can be
interpreted as the probability that an ad will be noticed when put in slot $j$ and it is
assumed that $\gamma_j > \gamma_{j+1}$ for all $1 \leq j \leq K$ and $\gamma_j = 0$ for $j > K$.
$e_i$ can be interpreted as the probability that an ad put by
bidder $i$ will be clicked on if noticed and is referred to as the {\it relevance}(quality score) of bidder
$i$. The
payoff/utility of bidder $i$ when given slot $j$ at a price of $p$ per-click is given by
$e_i\gamma_j (v_i - p)$. 
As of now, Google as well as Yahoo! use schemes closely modeled as RBR(rank by revenue)
with GSP(generalized second pricing). The bidders are ranked in the decreasing order of
$e_ib_i$ and the slots are allocated as per this ranks. Let the $\sigma(i)$ be 
the bidder allocated to the slot $i$ according to this ranking rule, then
$\sigma(i)$ is charged an amount equal to $\frac{e_{\sigma(i+1)} b_{\sigma(i+1)}}{e_{\sigma(i)}}$ per-click. This mechanism
has been extensively studied in recent years\cite{EOS05,Lah06,Var06,LP07,BCPP06}.
The solution concept that is widely adopted to study this auction game is a refinement of
Nash equilibrium called  {\it symmetric Nash equilibria(SNE)} 
independently proposed by Varian\cite{Var06} and Edelman et
al\cite{EOS05}. For notational simplicity, let $s_{\sigma(j+1)}= v_{\sigma(j+1)} e_{\sigma(j+1)}$ , then under this refinement, 
the revenue of the auctioneer at equilibrium is given by 
\begin{equation}
\sum_{i=1}^K  (\gamma_j - \gamma_{j+1}) j s_{\sigma(j+1)}.
\end{equation}
In this section, we discuss how to incorporate the capacity constraints 
in the keyword auctions being currently used by Google and Yahoo!.
We understand that there could be several ways for doing so, however, we consider a very simple and intuitive
way of incorporating the capacity constraints via probability spikes as follows:
\begin{itemize}
\item The first $K-1$ slots are sold as usual to the $K-1$ high-ranked bidders. 
\item The last slot (i.e. the $K$th slot) is sold via probability spikes $(p_1,p_2,\dots,p_M)$
among the $M$ bidders ranked $K$ through $K+M-1$. 
$M$ is chosen to accommodate as many more bidders as the auctioneer wants.
\item There is a single combined auction for both of the above. 

\end{itemize} 
Clearly, the single combined auction is equivalent to the keyword auction with $(K+M-1)$ slots with position based CTRs taken as
$(\gamma_1,\gamma_2,\dots,\gamma_{K-1}, \gamma_k p_1, \gamma_K p_2, \dots,\gamma_K p_M)$ i.e. $\tilde{\gamma}_i=\gamma_i$ for $i \leq K-1$,
$\tilde{\gamma}_{K+j-1} = \gamma_K p_j$ for $1 \leq j \leq M$ and $\tilde{\gamma}_i=0$ otherwise. 
The revenue of the auctioneer at {\em SNE} is  
\begin{displaymath}
\begin{array}{ll}
R  =  \sum_{j=1}^{K+M-1} (\tilde{\gamma}_j - \tilde{\gamma}_{j+1}) j s_{\sigma(j+1)} \\
   =   \sum_{j=1}^{K-2} (\gamma_j - \gamma_{j+1}) j s_{\sigma(j+1)} + \gamma_{K-1} (K-1) s_{\sigma(K)} 
  - \gamma_K p_1 (K-1) s_{\sigma(K)}\\
 + \gamma_K \sum_{j=K}^{K+M-1} (p_{j+1-K} - p_{j+2-K}) j s_{\sigma(j+1)}. \\
\end{array}
\end{displaymath}
Now, maximizing $R$ as a function of $p_j$'s is equivalent to maximizing the function 
\begin{displaymath}
\begin{array}{ll}
H & = - \gamma_K p_1 (K-1) s_{\sigma(K)} + \gamma_K \sum_{j=K}^{K+M-1} (p_{j+1-K} - p_{j+2-K}) j s_{\sigma(j+1)} \\
 & = \sum_{j=1}^M \theta_j \left\{(K+j-1) s_{\sigma(K+j)} - (K-1) s_{\sigma(K)}\right\}  \textrm{  where  }  \theta_j = p_j - p_{j+1} ; j=1,2,\dots,M-1 \& \theta_M = p_M \\
 & = \sum_{j=1}^M \theta_j j d_j \textrm{  where  } d_j = \frac{1}{j} \left\{(K+j-1) s_{\sigma(K+j)} - (K-1) s_{\sigma(K)}\right\}.
\end{array}
\end{displaymath}
Note that $H$ may not be gap-wise monotone, however, 
it is not hard to see that the similar linear programming analysis as in the Section \ref{genopt} can be done to 
compute the price of capacity in the present scenario of keyword auctions as well. We omit the details. 

\paragraph{Selling clicks via auctioneer-controlled probability spikes: a new model for sponsored search advertising:} 
The design framework in Section \ref{single},
suggests a  new model for SSA, where the clicks are sold directly and not indirectly via
allocating impressions. In the usual pay-per-click model, a click is assigned to one of
the advertisers who has been allocated an impression in the page currently being viewed by
the user; thus, the user's collective experience determines the probability that an
impression winner would be the beneficiary of the click, and having set up the impressions
and the user experience, the auctioneer does not actively control the probability with
which a click will be allocated to a bidder. In the new model, a click could be considered as the
indivisible item being sold via probability spikes. Instead of putting ads directly in the
slots, the auctioneer could put some categories/information related to the specific
keyword as a link. An auction based on probability spikes is run whenever a user clicks on
this link and the user is directly taken to the landing page of the winning advertiser.     

\section{Future Work}
\label{conc}
In the present paper, our main goal was to motivate {\it capacity} as a fundamental metric in designing auctions in the 
information economy and then to initiate study of such a design framework for some simple and interesting scenarios such as 
single indivisible item and sponsored search advertising. However, there are myriad of other interesting scenarios 
where the capacity-enabled framework should be interesting to study. For example, 
the auctions of digital goods\cite{GHW01,GHKSW06}, combinatorial auctions \cite{CSS06} for selling information goods in bundles, 
double auctions and ad-exchanges\cite{KP03} etc. Further, probably the most important question that remains to be addressed
is to identify the best way of putting capacity constraints and to see how generic the approach via 
probability spikes could be made.   
We can consider the most general case of selling any set of items e.g. heterogeneous, homogeneous, indivisible, or divisible or any combination their of. 
Let $\Sigma = \{\sigma_i\}$ be the set of all possible allocations. Further, the elements of $\Sigma$ are named such that 
$\sigma_i \succeq \sigma_{i+1}$, where $ \succeq$ is auctioneer's preference over allocations.
Then, these set of items can be sold with capacity constraints via probability spikes $p_i$'s with $p_i \geq p_{i+1}$,
 wherein 
$\sigma_i$ is enforced with probability $p_i$.  

\subsection*{Acknowledgements}
The authors thank Dork Alahydoian, Sushil Bikhchandani, Gunes Ercal, Himawan Gunadhi and Adam Meyerson for insightful discussions. 
The work of S.K.S was partially supported by his internship at NetSeer Inc., Los Angeles.


\begin{thebibliography}{100}

\bibitem{AGM06} G. Aggarwal, A. Goel, R. Motwani, Truthful Auctions for Pricing Search Keywords, EC 2006.
\bibitem{infoeco} U. Birchler and M. Butler, Information Economics, Routledge, 2007. 
\bibitem{BCPP06} T. Borgers, I. Cox,  M. Pesendorfer, V. Petricek, 
Equilibrium Bids in Sponsored Search Auctions: Theory and Evidence, Technical report, University of Michigan (2007).
\bibitem{BO06} S. Bikhchandani and J. M. Ostroy, From the assignment model to combinatorial auctions. In Combinatorial Auctions MIT Press 2006.
\bibitem{CSS06} P. Crampton, Y. Shoham, and R. Steinberg, Combinatorial Auctions, MIT Press, 2006.  
\bibitem{DGS86} Gabrielle Demange, David Gale, and Marilda Sotomayor, Multi-Item Auctions, Jour. Political Economy, 94, 863-872, 1986.
\bibitem{EOS05} B. Edelman, M. Ostrovsky,  M. Schwarz,
Internet Advertising and the Generalized Second Price Auction: Selling Billions of Dollars Worth of Keywords, American Economic Review 2007.
\bibitem{GHW01} A. V. Goldberg, J. D. Hartline, A. Wright, Competitive auctions and digital goods, SODA 2001.
\bibitem{GHKSW06} A. V. Goldberg, J. D. Hartline, A. R. Karlin, M. Saks, A. Wright, Competitive auctions, 
Games and Economic Behavior, Vol. 55, pp:242-269, 2006.  
\bibitem{krishna} V. Krishna, Auction Theory, Academic Press, San Diego, 2002. 
\bibitem{KP03} J. Kalagnanam and D. Parkes,  Auctions, Bidding and Exchange Design. In: Simchi-Levi, Wu, Shen: Supply Chain Analysis in the eBusiness Area, Kluwer Academic Publishers, 2003.
\bibitem{Lah06} S. Lahaie, An Analysis of Alternative Slot Auction Designs for Sponsored Search, EC 2006.
\bibitem{LP07} S.  Lahaie, D. Pennock, Revenue Analysis of a Family of Ranking Rules for Keyword Auctions, EC 2007.
\bibitem{milgrom} P. Milgrom, Putting Auction Theory to Work, Cambridge  University Press, 2004.
\bibitem{MSVV05} A. Mehta, A.   Saberi, U.  Vazirani, V.   Vazirani, AdWords and generalized on-line matching, FOCS 2005.
\bibitem{Nelson70}P. Nelson, Information and Consumer Behaviour, The Journal of Political Economy, vol 78, pp:311-329, 1970.  
\bibitem{PO06} S. Pandey, and C. Olston, Handling advertisements of unknown quality in search advertising, NIPS 2006.
\bibitem{SBRR07} S. K. Singh, V. P. Roychowdhury, M. Bradonji\'c, and B. A. Rezaei,
Exploration via design and the cost of uncertainty in keyword auctions, preprint (available at http://arxiv.org/abs/0707.1053).
\bibitem{cap-med} S. K. Singh, V. P. Roychowdhury, H. Gunadhi, and B. A. Rezaei, Capacity constraints and the inevitability 
of mediators in adword auctions, WINE 2007 (to appear).
\bibitem{diverse} S. K. Singh, V. P. Roychowdhury, H. Gunadhi, and B. A. Rezaei, Diversification in the Internet Economy: 
The Role of For-Profit Mediators, preprint (available at http://arxiv.org/abs/0711.0259).
\bibitem{SS72} L. S. Shapley and M. Shubik, The Assignment Game I: The Core, Int. J. Game Theory 1, no. 2, 111-30, 1972.
\bibitem{inforules} C. Shapiro and H. R. Varian, Information Rules: A Strategic Guide to the Network Economy, Harvard Business School Press, 1998.  
\bibitem{TK91} A. Tversky; D. Kahneman, Loss Aversion in Riskless Choice: A Reference-Dependent Model,
The Quarterly Journal of Economics, Vol. 106, No. 4. (Nov., 1991), pp. 1039-1061.
\bibitem{Var06} H. Varian, Position Auctions, To appear in International Journal of Industrial Organization.
\bibitem{WVLL07} J. Wortman, Y. Vorobeychik, L. Li, and J. Langford,
Maintaining equilibria during exploration in sponsored search auctions, WINE 2007 (to appear).

\end{thebibliography}
\end{document}